# Bagged Empirical Null $p$-values: A Method to Account for Model Uncertainty in Large Scale Inference

Sarah Fletcher Mercaldo*, Jeffrey D. Blume, Ph.D.

*Department of Biostatistics, Vanderbilt University, Nashville, TN*

sarah.fletcher@vanderbilt.edu

### SUMMARY

When conducting large scale inference, such as genome-wide association studies or image analysis, nominal $p$-values are often adjusted to improve control over the family-wise error rate (FWER). When the majority of tests are null, procedures controlling the False discovery rate (Fdr) can be improved by replacing the theoretical global null with its empirical estimate. However, these other adjustment procedures remain sensitive to the working model assumption. Here we propose two key ideas to improve inference in this space. First, we propose $p$-values that are standardized to the empirical null distribution (instead of the theoretical null). Second, we propose model averaging $p$-values by bootstrap aggregation (Bagging) to account for model uncertainty and selection procedures. The combination of these two key ideas yields *bagged empirical null $p$-values (BEN $p$-values)* that often dramatically alter the rank ordering of significant findings. Moreover, we find that a multidimensional selection criteria based on BEN $p$-values and bagged model fit statistics is more likely to yield reproducible findings. A re-analysis of the famous Golub Leukemia data is presented to illustrate these ideas. We uncovered new findings in these data, not detected previously, that are backed by published bench work pre-dating the Gloub experiment. A *pseudo-simulation* using the leukemia data is also presented to explore the stability of this approach under broader conditions, and illustrates the superiority of the BEN $p$-values compared to the other approaches.

---

*To whom correspondence should be addressed.

## 1. Introduction

Modern day technology like microarrays, RNA-sequencing, and fMRI Imaging, has given rise to a new era of statistical methods for high-throughput science. These methods are commonly referred to as large-scale inference, and can produce data corresponding to millions of statistical hypotheses for simultaneous testing. However, the repeated application of classical hypothesis testing methods can lead to concerns regarding control over the inflated Family-wise Error Rate (FWER), global Type I Error rate, and reduced statistical power (Efron (2012)).

Bonferroni adjustments, which provides strict control of the FWER, are highly conservative in the large-scale context and they are often avoided because of the dramatic loss of power that is associated with their use (Sham and Purcell (2014)). The literature is turning to other $p$-values adjustments, such as the Benjamini-Hochberg (B-H) False discovery rate (Fdr) procedure (Benjamini and Hochberg (1995)), which improves power by relaxing control of the FWER. The FWER can remain inflated, so long as the Fdr remains below a pre-specified threshold. This yields increased power because the FWER inflation is reduced, but not eliminated. Note that it is not possible to control both the FWER and Fdr simultaneously. Both quantities are functions of the per comparison Type I Error rate, so a compromise between must be made.

When the majority of tests are null and Fdr adjustments are warranted, Efron (2012) ad-vocated for replacing the theoretical null with its empirical estimate to improve the operating characteristics of Fdr procedures. This *empirical Fdr* procedure is an empirical Bayes procedure where the empirical null distribution is estimated from the mixture distribution of null and non-null test statistics. A more desirable error-rate tradeoff results because the mixture distribution of test statistics does not necessarily follow the theoretical null distribution in large-scale data, often because of a complex correlation structure.

While the use of the empirical null distribution has so far been limited to Fdr computations, this idea is readily propagated to the computation of $p$-values in large-scale data. Specifically, we propose standardizing $p$-value or test statistics of interest to the empirical null distribution. Despite the well documented problems with $p$-values, inference based on them is still more intuitive to applied scientists than inference based on false discovery rates. This approach has the added advantage of limiting the confusion between the inferential roles of Fdr and the $p$-value. The Fdr measures the tendency of the observed results to be misleading, while the $p$-value measures the degree to which the data are compatible with the null hypothesis (Blume (2011)). The former measures the uncertainty of the observed findings, while the latter is the metric of the strength of statistical evidence. We found through our investigations that empirically standardized $p$-values result in a desirable error-rate compromise between Bonferroni and Fdr methods.

We also observed extreme dependence of $p$-value and Fdr inference on working model assumptions. A common example is the assumption of a simple or trivially adjusted regression model in GWAS studies. In many cases, the sensitivity of the findings to this assumption was much higher than it was to the choice of the significance cutoff. To address this issue, we propose accounting for model uncertainty by model averaging via bootstrap aggregation (Bagging) when computing $p$-values or Fdrs. The resulting bagged empirical null (BEN) $p$-values almost always dramatically altered the rank ordering of significant findings. We also examined simple procedures that selecting findings on the basis of both BEN $p$-values and bagged model fit statistics. The idea is to favor significant findings from the model that fits the data better. Not surprisingly, this tended to yield reproducible findings more often because the stability of the inferential model is accounted for.

A potential downside to this approach is the additional computation time and knowledge base needed to bag models and compute the empirical null distribution. However, the changes in the $p$-value ranking were so dramatic in our examples and simulations, that we suggest this step never be skipped. An analysis of the famous Golub Leukemia data, using our proposed BEN $p$-value approach to rank potential genes of interest, leads us to new discoveries that were confirmed in animal models that had been published before the Golub *and others* (1999) study was implemented. This provides some external biological confirmation for the adequacy of our new methods.

### 1.1 *Background*

Large-scale data is often characterized by the simultaneous acquisition of data on many variables or endpoints, in such a manner that the number of observations $n$ is often a magnitude or more less than the number of endpoints $N$. Genome wide association studies (GWAS), functional magnetic resonance imaging (fMRI), and Tandem mass spectrometry (MS/MS) are some examples of scientific processes that may yield such large data sets. While the science and relevance of these large-scale endeavors has evolved over the last decade, the popular statistical approaches for handling simultaneous testing of many hypothesis - massive multiple comparison adjustments - has not seen similar growth.

The Bonferroni procedure, popularized in the statistical literature by Dunn (1959), remains the popular choice in many genetic sub-specialties largely due to its simplicity. For example, the commonly used Bonferroni threshold of $5 \times 10^{-8}$ will control the FWER to 0.05 if 1 million gene hypotheses are tested simultaneously (Pe'er *and others* (2008)). However, this adjustment is associated with a massive loss in statistical power. The Bonferroni procedure was never intended for these large-scale scenarios where individual hypotheses are of interest (Perneger (1998); Miller (2012)).

Competitors to the Bonferroni procedure have been considered, with some yielding substantial power gains in small-scale settings (Wright (1992)). As an alternative, Benjamini and Hochberg (1995) and separately Shaffer (1995) proposed relaxing control of the FWER and instead controlling the two-tailed Global False discovery rate (Remark A of the Supplement elaborates). Of course, the ranking and selection of top *p*-values for the purpose of identifying findings as scientifically significant and worthy of further study has seen much debate, e.g. see Sham and Purcell (2014).

Storey (2002) proposed a Bayesian interpretation of the Fdr, and Efron (2012) extended this work by proposing an empirical Bayes generalization of the Benjamini and Hochberg procedure: the Empirical Fdr and empirical local Fdr (fdr). The empirical Bayes approach relies on the assumption that a large majority of the individual null hypotheses are in fact null. As a result, the mixture distribution of observed test statistics can be well estimated using a classic empirical Bayes argument. An advantageous property of the Fdr is that the procedure is scalable as a function of the number of tests. Unlike other multiple-testing adjustments the Fdr procedure is not a test of a composite null hypothesis against a single alternative, and therefore can identify individual tests as significant (Mark E Glickman *and others* (2015)).

Importantly, nearly all adjustment procedures including B-H control of the Fdr or control of the two-tailed Global Fdr under the theoretical null, do not alter the *p*-value ranks; instead they simply adjusts the level at which a given *p*-value is considered "significant. The lone exception is the local false discovery rate, and this is due to the lack of smoothness of the empirical density estimator. The failure of these methods to change the rankings of signifiant endpoints raises the question of whether these procedures have different discrimination capability or are simply re-calibrations of each other. As we will see, the bagging and use of empirical null $p$-values does alter the rankings while maintaining error-rate control, which is an exciting advance.

## 1.2   *Organization of Paper*

The main idea is to use bagged empirical null $p$-values and bagged model fit criteria to identify interesting and statistically significant findings in large-scale contexts. We compare this new approach to the popular approaches in use today. The proposed methods are applied to a Leukemia microarray data set (Golub *and others* (1999)), studied to select genes which are differentially expressed in Acute Myeloid Leukemia (AML) versus Acute Lymphoblastic Leukemia (ALL), and BEN $p$-values combined with bagged model fit statistics are compared to current methods. Section 2 describes the leukemia gene expression data, gives an overview of the proposed Bagged Empirical Null (BEN) $p$-values and Fdr comparators, and provides a detailed algorithm for calculation of BEN $p$-values. Section 3 describes an innovative pseudo-simulation used to assess the proposed methods, and applies the BEN algorithm to the leukemia data. Section 4 presents the results and compares the performance of commonly used large-scale inference analysis techniques. Section 5 discusses our findings from the case study and provides some concluding remarks. Section 6 provides a brief overview of the Supplementary Materials available online.

## 2. Materials and Methods

Here we define and describe the computational algorithm for our novel Bagged Empirical Null (BEN) $p$-values, establish simulations to examine its performance and apply it to the famous Gloub leukemia data set (Golub *and others* (1999)).

### 2.1    *Leukemia data*

The publicly available leukemia data consists of gene expression data for classification of leukemia into two types, Acute Lymphoblastic Leukemia (ALL) and Acute Myeloid Leukemia (AML), the latter of which has the worse prognosis. Each type of leukemia responds to different chemotherapies, so correct classification is important to a patients' treatment success. The leukemia data consist of 72 patients, 47 ALL cases and 25 AML cases all genotyped using Affymetrix Hgu6800 chips, resulting in 7129 gene expressions. The most extreme value was excluded, resulting in 7128 gene expressions used to identify interesting genes whose expression levels differ between ALL and AML subjects. Additional covariate information for patients, such as dichotomized age (children vs. adults), drawn sample location (peripheral blood vs. bone marrow), and gender (males vs. females) are available for analyses. The raw data are publicly available online via the 'golubEsets' package in Bioconductor. Gender was missing for 23 patients, and for each simulation missing covariate information was imputed using single imputation and predictive mean matching. As described by Efron (2012), data were normalized in order to eliminate response disparities among microarrays and reduce the impact of outlying values.

### 2.2    *The Empirical Null*

$P$-values and Fdr computations rely on the assumption that the theoretical null density is known. However, when conducting genome wide studies, this is a strong assumption that is likely false for a significant number of genes or SNPs. Nevertheless, it is highly believable that a very large proportion are null, leading to a situation where the null mixing distribution can be well estimated from the data at hand. As described by Efron (2012), there are several reasons the theoretical null distribution may fail. (1) Failed mathematical assumptions: It is possible that the test statistics are identically distributed as $N(0,1)$ but not independent, which means the mixing distribution is no longer $N(0,1)$. (2) Correlation across sampling units: Minor experimental defects can manifest themselves as correlation across sampling units. (3) Correlation across cases: Independence among the genes is not needed for valid false discovery rate inference only if we are using the correct null distribution. (4) Unobserved covariates: Additional confounding variables, not accounted for using naive approaches, are perhaps the most common reason why the theoretical null may fail, making it critical to account for model uncertainty.

For an individual gene $i$, the usual null hypothesis is $H_{0i}$: gene $i$ is null. The corresponding $z$-statistic, under normal assumptions, follows a standard normal distribution such that $H_{0i}$: $z_i \sim N(0, 1)$. However, by examining the empirical mixing distribution of $z_i$'s, we can assess if the typical theoretical null is supported by the data. We differentiate the individual null densities, $f_0(z_i)$, and non-null densities, $f_1(z_i)$, for $i = 1 \ldots N$ genes. We denote the mixture null comprised of all the $z$-values, $f(z) = \pi_0 f_0(z) + \pi_1 f_1(z)$ where $\pi_0$ and $\pi_1$ are the null and non-null prior probabilities (Efron (2012)).

The empirical null can be estimated by central matching or maximum likelihood estimation, both of which rely on the assumption that some central region of the empirical distribution exists where all genes are not differentially expressed (Remark B of the Supplement for details). Empirical null methods essentially accommodate a "blurry" null hypotheses, in which the uninteresting cases can deviate in minor ways from the theoretical null formulation.

Note that variations of the empirical null are possible depending on the desired modeling assumptions. For example, one can assume normality and estimate both the mean and the variance. Or, alternatively, one could fix the mean at zero and estimate the variance. The latter is intriguing, as it seems natural that empirical null should also be zero centered because the correlation would not affect centering. However, we explored variations like these and found little difference when compared to the classic empirical null algorithms, so the results are not included here.

### 2.3 *Model Selection and Bootstrap Aggregation*

Misspecification of the underlying model can be a real problem for reliable inference. For example, when testing for differentially expressed genes, a common practice is to perform $N$ $t$-tests, where $N$ is equal to the number of genes. This is equivalent to $N$ univariate linear models, each regressing gene expression on disease status. Often, a logistic regression model is more appropriate, since the goal is to use gene expression to predict leukemia status. In addition, the set of models considered ought to be adjusted for potentially confounding effects, allow for nonlinear effects, and possibly include robust standard errors. All of these things will significantly impact the final inference, but they are often ignored in routine applications.

A way to address many of these concerns is to use bootstrap aggregation (bagging) Breiman (1996). This method averages estimands of interest from a set of bootstrapped models. Bagging is often applied to situations where the estimator heavily depends on the sample, and perturbations of the sample may lead to significant changes in the statistical measure. Bagging has been shown to improve biological inference on gene sets (Jaffe *and others* (2013)), however the usefulness of this method has only been investigated for estimating the probability that a significant gene finding will replicate, so our application here is novel.

### 2.4 *Algorithm*

The algorithm for computing BEN $p$-values consists of the following steps:

1. Choose the set of generalized linear models appropriate for the statistical question; including potential covariate adjustments, interactions, non-linear effects, and robust standard errors.

2. Resample the data $B$ times, and fit the entire set of models from Step 1 in each of the $B$ bootstrap resamples, for each of the $N$ genes being tested.

3. Obtain the model fit statistic (e.g., AUC or $R^2$), and the z-statistic corresponding to the

hypothesis test of interest (i.e., z-value corresponding to, say, gene expression in the model).

4. Use the $N$ z-values from each model to estimate the Empirical Null distribution, $N(\mu, \sigma)$.

5. Standardize the test statistics to the empirical null distribution. Compute $p$-values and and Fdrs under the empirical null as usual.

6. For each of the N endpoints, select the model with the 'best' model fit (e.g., best AIC) and save all statistics of interest for that model.

7. Repeat $B$ times.

8. Take the average of the $B$ statistics for each of the $N$ genes.

9. Report the average empirical null $p$-value and averaged model fit statistic (e.g., AUC). These are the bagged statistics.

10. Flag genes as significant if they meet a multi-dimensional threshold on the BEN $p$-value and bagged fit statistic.

2.4.1 *Step 3: Calculating z-values for each variable of interest from the p-values associated with the gene covariate in each model* Calculate the $p$-value in the usual way associated with the gene covariate of interest from each model. From the $p$-value , calculate $z$-values such that $z_i = \Phi(p)$ for $i = 1, \ldots, N$ genes. If genes are allowed to have nonlinear associations, and/or interactions, use the $p$-value from the chunk test, and then transform back to $z$-scale in the same manner.

2.4.2 *Details on Step 4: Estimating the Empirical Null* There are several ways to 'bag' the empirical null distribution. The most principled approach is described in the algorithm above, where each model is fit, and an empirical null distribution is estimated from the $z$ values obtained from a single model. If there are $m$ models to choose from, $m$ empirical nulls will be estimated, and $m*N$ statistics, will be calculated in total. We call this the principled approach because the underling working model is held fixed across the genes. Then, for each of the $N$ genes, the set of desired statistics is chosen from the model with the 'Best fit' (e.g., lowest AIC). An alternative is this: after the $m$ models are fit, the $z$-values are selected from the 'Best fit' model, and then a *single* empirical null is estimated from the $N$ $z$-values. Here the working model is allowed to vary across the genes. Lastly, one could estimate a single empirical null distribution from the combined $m*N$ $z$-values across all models. This approach adds a layer of correlation to the mixture distribution but is not favoring the 'best' fit model. These last two approaches to computing the empirical null distribution have the advantage of only one computation cycle, speeding up the algorithm. There was little difference between the empirical null procedures in practice, and we found that other approaches did not significantly alter the BEN $p$-value rankings. For this paper, we used the "principled" approach, and the two alternative methods are explored in Remark C of the Supplement since they appear to be viable shortcuts that merit further exploration.

## 3. Simulation and Implementation

### 3.1 *Pseudo-Simulation*

Large-scale data is often defined by its rich and complex correlation structure, which often extends non-uniformly over columns, rows, and clusters. As such, it is very hard to simulate realistic large-scale data from scratch. Because of this, we took a "Pseudo-Simulation" approach. Our idea is to take the subset of Gloub null genes and remove any mean effects via a highly parameterized regression including several covariates and interactions. The matrix of residuals retains the complex correlation structure, upon which we add back fixed effects via an assumed regression model (this is our simulation engine). We refer to this simulation as 'Pseudo' because we are not generating *new* error structures. In effect, we get a simulation using a real-world error structure and thus retain the complexity of these large-scale data that is often critical to the evaluation of novel methods.

Specifically, we did the following: In the original unchanged leukemia data, fit univariate linear models for each gene, and then took the subset of genes whose $p$-value associated with the AML/ALL coefficient was greater than 0.3. This resulted in 3172 genes whose gene expressions we assumed have little to no association with leukemia type. We then randomly selected 30 genes within this subset, computed the residuals after regressing over as many effects as we could, and induced 10 genes to have a strong association, 10 genes to have a moderate association, and 10 genes to have a weak association between leukemia type and gene expression.

To selectively adjust certain gene expression levels and leukemia relationships, new expression data was estimated by adding new fitted values,

$$\{\texttt{Gene Expression}^*\}_j = \{\boldsymbol{X\beta}^*\}_j$$

$$\begin{aligned} = \beta_{0j}^* + \beta_{1j}^*\{\texttt{Leukemia Type}\} + \beta_{2j}^*\{\texttt{Gender}\} + \beta_{3j}^*\{\texttt{Sample}\} \\ + \beta_{4j}^*\{\texttt{Leukemia Type x Gender}\} + \beta_{5j}^*\{\texttt{Leukemia Type x Sample}\} \\ + \beta_{6j}^*\{\texttt{Gender x Sample}\} \end{aligned}$$

to the original residuals, $\hat{\epsilon}_j = \{\texttt{Gene Expression}\}_j - \{\boldsymbol{X\hat{\beta}}\}_j$, for each of the induced genes $j = 1 \ldots 30$. This results in $\{\texttt{Gene Expression}^{\text{NEW}}\}_j = \{\texttt{Gene Expression}^*\}_j + \hat{\epsilon}_j$.

This allows us to specify any level of complexity in the model, while preserving the original error structure from the data. Each set of 10 strong, moderate, and weak genes comprised of 2 genes induced to be associated with just leukemia type ($\beta_1^*$), 2 genes induced to have a gender and a leukemia type by gender interaction ($\beta_2^*, \beta_4^*$), 2 genes induced to have a leukemia type and a leukemia type by gender interaction ($\beta_1^*, \beta_4^*$), 2 genes induced to have a leukemia type and a leukemia type by sample interaction ($\beta_1^*, \beta_5^*$), and 2 genes induced to have a leukemia type, leukemia type by gender, and a leukemia type by sample interaction ($\beta_1^*, \beta_4^*, \beta_5^*$). Strong gene associations, moderate gene associations, and weak gene associations were induced to have $\boldsymbol{\beta}^*$ values 7, 4, and 2 times the original leukemia type coefficients respectively. These values were chosen such that the multiplicative effect of the coefficient outweighed their corresponding empirical standard errors.

The pseudo-simulated data was then normalized, and the BEN algorithm was implemented. The pseudo-simulations were conducted under linear regression models because (1) they provide a direction comparison to methods based on single univariate $t$-statistics, common in the literature and (2) their residuals are well defined when compared to logistic regression. We expect the results would extend to logistic regression, and our primary example uses logistic regression. The following linear models were considered for the BEN algorithm:

(1) $E[\texttt{Gene Expression}] = \beta_0 + \beta_1\{\texttt{Leukemia Type}\}$

(2) $E[\texttt{Gene Expression}] = \beta_0 + \beta_1\{\texttt{Leukemia Type}\} + \beta_2\{\texttt{Sample}\}$

(3) $E[\texttt{Gene Expression}] = \beta_0 + \beta_1\{\texttt{Leukemia Type}\} + \beta_2\{\texttt{Gender}\}$

(4) $E[\texttt{Gene Expression}] = \beta_0 + \beta_1\{\texttt{Leukemia Type}\} + \beta_2\{\texttt{Sample}\} + \beta_3\{\texttt{Gender}\}$

(5) $E[\texttt{Gene Expression}] = \beta_0 + \beta_1\{\texttt{Leukemia Type}\} + \beta_2\{\texttt{Sample}\} + \beta_3\{\texttt{Gender} + \beta_4\{\texttt{Leukemia Type x Sample}\}$

(6) $E[\texttt{Gene Expression}] = \beta_0 + \beta_1\{\texttt{Leukemia Type}\} + \beta_2\{\texttt{Sample}\} + \beta_3\{\texttt{Gender} + \beta_4\{\texttt{Leukemia Type x Gender}\}$

(7) $E[\texttt{Gene Expression}] = \beta_0 + \beta_1\{\texttt{Leukemia Type}\} + \beta_2\{\texttt{Sample}\} + \beta_3\{\texttt{Gender} + \beta_4\{\texttt{Sample x Gender}\}$

(8) $E[\texttt{Gene Expression}] = \beta_0 + \beta_1\{\texttt{Leukemia Type}\} + \beta_2\{\texttt{Sample}\} + \beta_3\{\texttt{Gender} + \beta_4\{\texttt{Leukemia Type x Gender}\}+$

$\beta_5\{\texttt{Leukemia Type x Sample}\}$

### 3.2   *Logistic Regression Models Applied to the Leukemia Data*

Two separate implementations of the BEN algorithm were performed for the leukemia data using both linear regression and logistic regression. Since the set of logistic models are more appropriate for answering the scientific question at hand, these are the models discussed here. The linear regression results are included in Remark D of the Supplement.

7128 regression models were fit, and the models included all combinations of the covariates gene expression (continuous), site of sample (peripheral blood vs. bone marrow), and gender (male vs. female). In two of the regression models, gene expression was modeled flexibly using restricted cubic splines, as gene expression does not necessarily have a linear relationship with leukemia type.

The following models were considered for each of the 500 bootstrap aggregations:

(1) $logit(E[\texttt{Leukemia Type}]) = \beta_0 + \beta_1\{\texttt{Gene Expression}\}$

(2) $logit(E[\texttt{Leukemia Type}]) = \beta_0 + \beta_1\{\texttt{Gene Expression}\} + \beta_2\{\texttt{Sample}\}$

(3) $logit(E[\texttt{Leukemia Type}]) = \beta_0 + \beta_1\{\texttt{Gene Expression}\} + \beta_2\{\texttt{Gender}\}$

(4) $logit(E[\texttt{Leukemia Type}]) = \beta_0 + \beta_1\{\texttt{Gene Expression}\} + \beta_2\{\texttt{Sample}\} + \beta_3\{\texttt{Gender}\}$

(5) $logit(E[\texttt{Leukemia Type}]) = \beta_0 + \beta_{1,2}\{rcs(\{\texttt{Gene Expression}\}, 3)\}$

(6) $logit(E[\texttt{Leukemia Type}]) = \beta_0 + \beta_{1,2}\{rcs(\{\texttt{Gene Expression}\}, 3)\} + \beta_3\{\texttt{Sample}\} + \beta_4\{\texttt{Gender}\}$

For models (1) through (4) the test of $\beta_1$ provides $N = 7128$ Wald $z$-statistics used in the BEN algorithm. For models (5) and (6) the $p$-value from the chunk test of $\beta_1 = \beta_2 = 0$, the terms associated with the non-linear effect of gene expression, is transformed back to the $z$-statistic as described in 2.4.1. In order to compare results to more traditional methods, univariate logistic regression models were fit, and the $N = 7128$ $z$-values were used to calculate $p$-values and empirical null $p$-values. All $p$-values were adjusted using the Bonferroni and Benjamini-Hochberg procedures for comparison.

## 4. Results

### 4.1 *Pseudo-Simulation*

Univariate linear regression models were fit for every gene using the manipulated data: $E\{\texttt{Gene Expression}\}_i = \beta_{0i} + \beta_{1i}\{\texttt{Leukemia Type}\}$ for $i = 1 \ldots 3172$. The unadjusted, Bonferroni, and Benjamini-Hochberg $p$-values, along with each model $R^2$ was computed. Table 1 presents the results of the pseudo-simulation after performing 250 simulations of 250 bootstrap aggregations.

Investigating all subsets of $p$-values $\leqslant 0.05$, the BEN $p$-values found more truly associated genes than the Bonferroni corrected $p$-values from the univariate models. This increase in power comes with the cost of a slight increase in empirical False Discovery Rate (FDR) for the BEN $p$-values compared to the Bonferroni corrected $p$-values. Interestingly, we estimated 115 genes from the univariate model to have $p$-values $\leqslant 0.05$, and only 16 of those had true disease status/gene expression relationships. Bagging $p$-values reduces the total number genes with $p$-values $\leqslant 0.05$ to approximately 105, with about 20 of those being true relationships. Our results indicate that bagging alone yields a desirable increase in power and decrease in FDR apart from the empirical null methodology. BEN $p$-values have a comparable power to unadjusted and empirical null $p$-values, but they result in a significantly reduced FDR.

When using the multi-dimensional metric, $p$-values $\leqslant 0.05$ and $R^2 \geqslant 0.5$ (from either the univariate models or the bagged models), the BEN $p$-values and Bagged $p$-values have the largest power and have similar FDR control. When selecting genes based on $p$-value alone, EN, unadjusted, and Bagged $p$-values have similar properties (with Bagged $p$-values having the greatest power), however when selecting genes sets based on this dual metric the usefulness of bagging and empirical null procedures is apparent.

It may be tempting to find the traditional Bonferroni $p$-values still favorable since the empirical FDR is low (especially when using both the $p$-value and model fit statistic), however the Bonferroni $p$-values mostly select the genes whose true model is univariate (i.e. 4 of the 6 models where Leukemia Type and Gene expression, $\beta_1^*$, was induced to have a relationship). So the Bonferroni adjustment might be considered ultra-conservative if the underling model is correctly specified, but if not the adjustment appears to overly penalize. All other models were not selected by the Bonferroni $p$-values. Bagging allows for the results to favor the 'best fit' model, which is more likely to be close to the correct but unknown model, and this results in more true discoveries.

| | True Discoveries | | | | False Discoveries | Power | FDR |
|---|---|---|---|---|---|---|---|
| | Strong | Moderate | Weak | All | Ordered ↓ Type I Error* | #True Dis./#Induced Genes | #False Dis./#Total Dis. |
| **$p$-value $\leqslant 0.05$** | | | | | | | |
| EN | 8 (7,8) | 6 (5,7) | 3 (2,4) | 17 (16,19) | 138 (136,139) | 0.57 (0.53,0.63) | 0.89 (0.88,0.9) |
| Unadjusted | 7 (6,8) | 6 (5,7) | 3 (2,4) | 16 (15,18) | 99 (99,100) | 0.53 (0.5,0.6) | 0.86 (0.85,0.87) |
| Bagged | 9 (8,10) | 8 (7,9) | 4 (3,5) | 21 (19,22) | 85 (74,101) | 0.67 (0.63,0.73) | 0.81 (0.78,0.84) |
| **BEN** | 8 (7,9) | 6 (5,7) | 2 (1,2) | 16 (14,17) | 12 (11,13) | 0.53 (0.47,0.57) | 0.43 (0.4,0.46) |
| Bagged B-H | 8 (7,9) | 6 (5,7) | 2 (1,2) | 16 (14,17) | 10 (10,13) | 0.53 (0.47,0.57) | 0.42 (0.39,0.46) |
| EN B-H | 5 (4,7) | 3 (2,4) | 0 (0,1) | 8 (8,11) | 8 (8,8) | 0.3 (0.27,0.37) | 0.47 (0.42,0.5) |
| B-H | 5 (4,6) | 3 (2,4) | 0 (0,1) | 8 (7,10) | 8 (8,8) | 0.3 (0.23,0.33) | 0.47 (0.44,0.53) |
| EN Bonferroni | 5 (4,6) | 2 (2,4) | 0 (0,1) | 7 (6,9) | 5 (5,5) | 0.23 (0.2,0.3) | 0.42 (0.36,0.45) |
| Bonferroni | 4 (3,5) | 2 (1,3) | 0 (0,0) | 6 (5,8) | 3 (3,3) | 0.23 (0.17,0.27) | 0.3 (0.27,0.38) |
| Bagged Bonferroni | 7 (6,8) | 4 (3,5) | 0 (0,0) | 11 (9,12) | 2 (1,2) | 0.37 (0.3,0.4) | 0.14 (0.09,0.17) |
| BEN B-H | 6 (5,7) | 2 (1,3) | 0 (0,0) | 8 (7,10) | 0 (0,0) | 0.27 (0.23,0.33) | 0 (0,0) |
| BEN Bonferroni | 5 (4,6) | 1 (0,2) | 0 (0,0) | 6 (5,7) | 0 (0,0) | 0.2 (0.17,0.23) | 0 (0,0) |
| **$p$-value $\leqslant 0.05$ & $R^2 \geqslant 0.5$** | | | | | | | |
| Bagged | 7 (6,8) | 4 (3,5) | 0 (0,1) | 11 (9,12) | 2 (1,2) | 0.37 (0.3,0.4) | 0.12 (0.08,0.15) |
| **BEN** | 7 (6,8) | 4 (3,5) | 0 (0,1) | 11 (9,12) | 1 (1,1) | 0.37 (0.3,0.4) | 0.08 (0.08,0.1) |
| Bagged B-H | 7 (6,8) | 4 (3,5) | 0 (0,1) | 11 (9,12) | 1 (1,1) | 0.37 (0.3,0.4) | 0.09 (0.08,0.11) |
| Bagged Bonferroni | 7 (6,8) | 3 (2,4) | 0 (0,0) | 10 (9,11) | 1 (1,1) | 0.33 (0.3,0.37) | 0.09 (0.08,0.1) |
| BEN B-H | 6 (5,7) | 2 (1,3) | 0 (0,0) | 8 (7,10) | 0 (0,0) | 0.27 (0.23,0.33) | 0 (0,0) |
| BEN Bonferroni | 5 (4,6) | 1 (0,2) | 0 (0,0) | 6 (5,7) | 0 (0,0) | 0.2 (0.17,0.23) | 0 (0,0) |
| EN | 2 (1,3) | 0 (0,1) | 0 (0,0) | 2 (1,3) | 0 (0,0) | 0.07 (0.03,0.1) | 0 (0,0) |
| Unadjusted | 2 (1,3) | 0 (0,1) | 0 (0,0) | 2 (1,3) | 0 (0,0) | 0.07 (0.03,0.1) | 0 (0,0) |
| EN B-H | 2 (1,3) | 0 (0,1) | 0 (0,0) | 2 (1,3) | 0 (0,0) | 0.07 (0.03,0.1) | 0 (0,0) |
| B-H | 2 (1,3) | 0 (0,1) | 0 (0,0) | 2 (1,3) | 0 (0,0) | 0.07 (0.03,0.1) | 0 (0,0) |
| EN Bonferroni | 2 (1,3) | 0 (0,1) | 0 (0,0) | 2 (1,3) | 0 (0,0) | 0.07 (0.03,0.1) | 0 (0,0) |
| Bonferroni | 2 (1,3) | 0 (0,1) | 0 (0,0) | 2 (1,3) | 0 (0,0) | 0.07 (0.03,0.1) | 0 (0,0) |

Table 1. Results of the Pseudo-simulation with 250 replications and 250 bootstrap aggregations. The Median and (IQR) are presented here. There are 3172 genes tested in total, with 10 genes induced to have a strong, 10 genes induced to have a moderate, and 10 genes induced to have a weak association between Gene Expression and Leukemia Type. True Discoveries are those genes found which had a true induced relationship between Gene Expression and Leukemia type. False Discoveries are those genes which had no induced relationship between gene expression and leukemia type. The FDR is the number of false discoveries out of all discovered genes. The Power is the number of true differentially expressed genes out of the 30 genes induced to have a true relationship between gene expression and leukemia type. *$p$-value type is ordered by descending Type I Error (Number of False discoveris/Truly Null Genes). BEN: Bagged Empirical Null, B-H: Benjamini-Hochberg, EN: Empirical Null.

Note that in the pseudo-simulation the Empirical Null (EN) $p$-values tended to have more false discoveries than the unadjusted $p$-values. We expect this to be true if we assume 3142 genes are truly null, then we would expect 157 false discoveries by chance alone, a result in line with the EN $p$-values. The combination of bagging and empirical null procedures is more likely to select the truly differentially expressed genes (Table 1). It also has a more desirable FDR/Power tradeoff, and therefore should be preferred over other $p$-value adjustment techniques.

### 4.2 *Application to Leukemia Data*

4.2.1 *Naive Approaches to Analysis of Leukemia Data* For comparison, the usual analyses were performed using the leukemia data. Simple Logistic regression models were fit, $logit(E[\texttt{Leukemia Type}_i]) = \beta_0 + \beta_1\{\texttt{Gene Expression}_i\}$ for $i = 1 \dots 7128$ genes. The conventional analysis is shown in Figure 1, where the unadjusted p-values, Bonferroni corrected $p$-values, and global two-tailed B-H (Fdr) $p$-values are presented, sorted from smallest to largest. The black dashed line represents the same $p$-values calculated under the empirical $N(0.13, 1.70)$ null using the maximum likelihood method (slight mean shift with large variance inflation). The solid blue line is the B-H $p$-values (the two-tailed global Fdrs) calculated under the theoretical null, and the solid red line is the Bonferroni corrected $p$-values. The EN $p$-values are an interesting compromise between the strict Bonferroni and more relaxed B-H $p$-value adjustments. For these data, there are 391 EN $p$-values $\leqslant 0.05$, compared to 7 Bonferroni $p$-values and 971 B-H $p$-values. EN $p$-values appear to self-adjust, providing a more desiravel error-rate tradeoff than Bonferroni or B-H $p$-values, without having to do post-hoc adjustments.

The bagged theoretical null $p$-values, and bagged theoretical null B-H $p$-values are shrunk towards 0.5, compared to the non-bagged counterparts (Figure 1: right plot). Since $p$-values follow a Uniform(0,1) distribution under the null hypothesis, the mean of the null $p$-values is approximately 0.5 (Remark E of the Supplement), so this behavior is expected. The display of bagged results is ordered by the original unadjusted $p$-value magnitude.

4.2.2 *Previous reports on differentially expressed genes for leukemia* There is no known list or compilation of differentially expressed genes related to the leukemia. A literature search yielded 5 independent and highly cited papers that present gene findings, specific to leukemia, based on their analyses. We reviewed and compared their findings and compared these results to the genes the BEN algorithm ultimately selected. Golub *and others* (1999); Lee *and others* (2003); Bø and

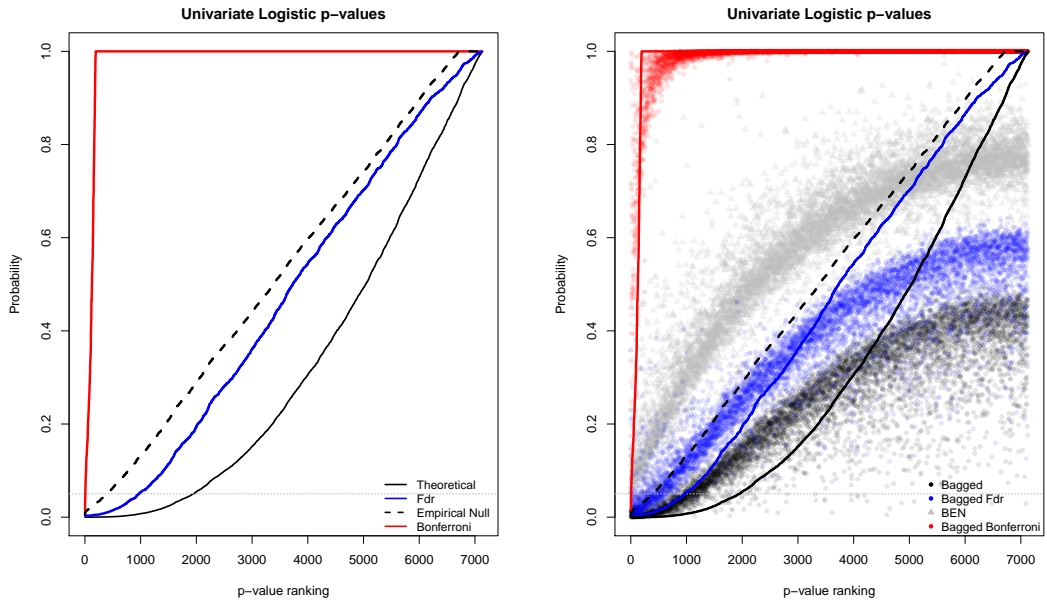

Fig. 1. Left Figure: The black solid line represents the unadjusted *p*-values associated with the gene expression coefficient from each of the 7128 univariate logistic regression models. The *p*-values are sorted from smallest to largest. The black dashed line represents the p-values recalibrated using the empirical $N(0.13, 1.70)$ null derived from the maximum likelihood method. The solid blue line represents The Benjamini-Hochberg (B-H) *p*-values, equivalent to the two-tailed global false discovery rate (Fdr) calculated under the theoretical null. The solid red line represents the Bonferroni corrected p-values under the theoretical null. Right Figure: The black dots are the bagged *p*-values calculated under the theoretical null. The gray triangles are the bagged empirical null (BEN) *p*-values. The blue dots are the bagged B-H *p*-values calculated under the theoretical null. The red dots are the bagged Bonferroni *p*-values calculated under the theoretical null. All points are sorted based on the original unadjusted *p*-values.

Jonassen (2002); Zhou *and others* (2004); Tong and Schierz (2011) use a variety of statistical methods to classify or predict gene expressions associated with leukemia type. Note there is a lot of variability in the reported genes between studies.

Golub *and others* (1999) reports 50 top genes (25 genes most differentially expressed in AML patients, and 25 genes most differentially expressed ALL patients). Lee *and others* (2003) reports the 27 genes which are the best classifiers of ALL vs. AML using a Bayesian variable selection approach. Bø and Jonassen (2002) report the top 50 genes for ALL/AML class separation using their all pairs subset selection procedure. Zhou *and others* (2004) report the top 20 important genes selected using their proposed Bayesian gene selection algorithm. Tong and Schierz (2011) report the 22 genes selected by their genetic algorithm-neural network. For comparison, we also compare the top 20 smallest univariate logistic regression *p*-value and the 7 Bonferroni adjusted *p*-values $\leqslant 0.05$ associated with leukemia type.

The BEN algorithm resulted in 22 genes who met the dual criterion of BEN *p*-value $\leqslant 0.1$ and bagged AUC $\geqslant 0.90$ (Table 2). Of this set, the gene with the smallest *p*-value is M83667 a known transcription factor of the NF-IL6-beta protein. A study by Natsuka *and others* (1992), that predates the leukemia study, found there was a drastic increase in expression of NF-IL6 messenger RNA (mRNA) during the differentiation to a macrophage lineage in mouse myeloid leukemia cells using mouse models. The original Golub *and others* (1999) paper does not report this gene as one of top 50 genes differentially expressed in AML/ALL patients. Lee *and others* (2003) and Zhou *and others* (2004) also fail to identify this gene. Given the prior prominence of M83667, which appears to be a major omission.

4.2.3 *Comparison of p-values* We can visualize the singular effects of just bagging *p*-values, of just recalibrating the *p*-values to the empirical null distribution, and the additive effect of bagging and empirical null procedures in Figure 2. Compared to the original univariate *p*-values from the univariate logistic regression models, the bagged *p*-values are shrunk towards 0.5. Those *p*-values which were most likely null had corresponding bagged *p*-values that were shrunk towards 0.5, and the original *p*-values which were closer to 0 had larger bagged *p*-values. Notice that the empirical null *p*-values are a one-to-one function of the original *p*-values, and incorporating the empirical null amounts to shifting the original *p*-values away from 0. The BEN *p*-values in Figure 2 show a left horizontal and vertical shift of the *p*-values compared to the original *p*-values. We investigated the effects of just bootstrapping the *p*-values and include those results in Remark F of the Supplement. We compared the top 50 *p*-values reported by Golub *and others* (1999), the 7 *p*-values with a Bonferroni adjustment $\leqslant 0.05$, and NFIL6beta protein mRNA, the top gene

| Gene | BEN p-value | AUC | Golub | Bo | Lee | Tong | Zhou | Univariate $p$ | Bonferroni $p$ |
|------|-------------|-----|-------|-----|-----|------|------|----------------|----------------|
| M83667 | 0.057 | 0.922 | | X | | X | | X | X |
| M33195 | 0.064 | 0.912 | | | | | | X | |
| U22376 | 0.066 | 0.904 | X | | X | | X | X | |
| X97267 | 0.069 | 0.908 | | | | X | | | |
| M62762 | 0.073 | 0.927 | X | | | | | | |
| X52056 | 0.075 | 0.913 | | X | | X | | | |
| X78669 | 0.076 | 0.902 | | | | | | | |
| M31211 | 0.076 | 0.927 | X | X | | | X | X | |
| M22960 | 0.077 | 0.922 | | | X | X | | | |
| M19507 | 0.078 | 0.932 | | X | X | X | | | |
| Z15115 | 0.080 | 0.954 | X | X | | | X | X | |
| U41635 | 0.082 | 0.902 | | | | | | | |
| D14664 | 0.082 | 0.910 | | | | | | | |
| L09717 | 0.082 | 0.909 | | | | | | | |
| X17042 | 0.084 | 0.910 | X | X | | | | | |
| U16954 | 0.084 | 0.922 | | X | | | | | |
| J03801 | 0.088 | 0.901 | | | | | | | |
| M32304 | 0.090 | 0.906 | | | | | | | |
| M93056 | 0.092 | 0.907 | | | | | | | |
| U77948 | 0.095 | 0.911 | | X | | | | | |
| U57721 | 0.097 | 0.913 | | | | | | | |
| X07743 | 0.098 | 0.917 | | | X | | | | |

Table 2. The 22 leukemia data genes with bagged empirical null (BEN) $p$-values $\leqslant 0.1$ and bagged AUC $\geqslant 0.9$. Genes are sorted by (BEN) $p$-values from smallest to largest. We reviewed Golub *and others* (1999); Lee *and others* (2003); Bø and Jonassen (2002); Zhou *and others* (2004); Tong and Schierz (2011), and an X is placed where the authors reported a gene we also selected. The univariate $p$-values are derived from the univariate logistic regressions of the original data, and the resulting top 20 $p$-values are compared to the BEN results. The Bonferroni correction was applied to the univariate model $p$-values, and all Bonferroni $p$-values $\leqslant 0.05$ are compared to the BEN results.

selected by the BEN algorithm 2-dimensional criterion in Table 2.

### 4.2.4 *Visualizing the 2-dimensional criteria*  We propose using a dual criterion of $p$-values and model fit statistics (AUC) for selecting a subset of genes for further investigation (Figure 3). Visualizing the relationship between $p$-values and AUC from the leukemia data makes it clear that even small $p$-values can have a low corresponding AUC. The AUC cutoff criterion (horizontal dashed lines) could be adjusted up or down to further restrict or grow the set of genes for study. We have chosen 0.9, but this threshold (or the similar $R^2$ threshold in Table 1) could be selected specifically to choose a subset whose size is feasible for further study.

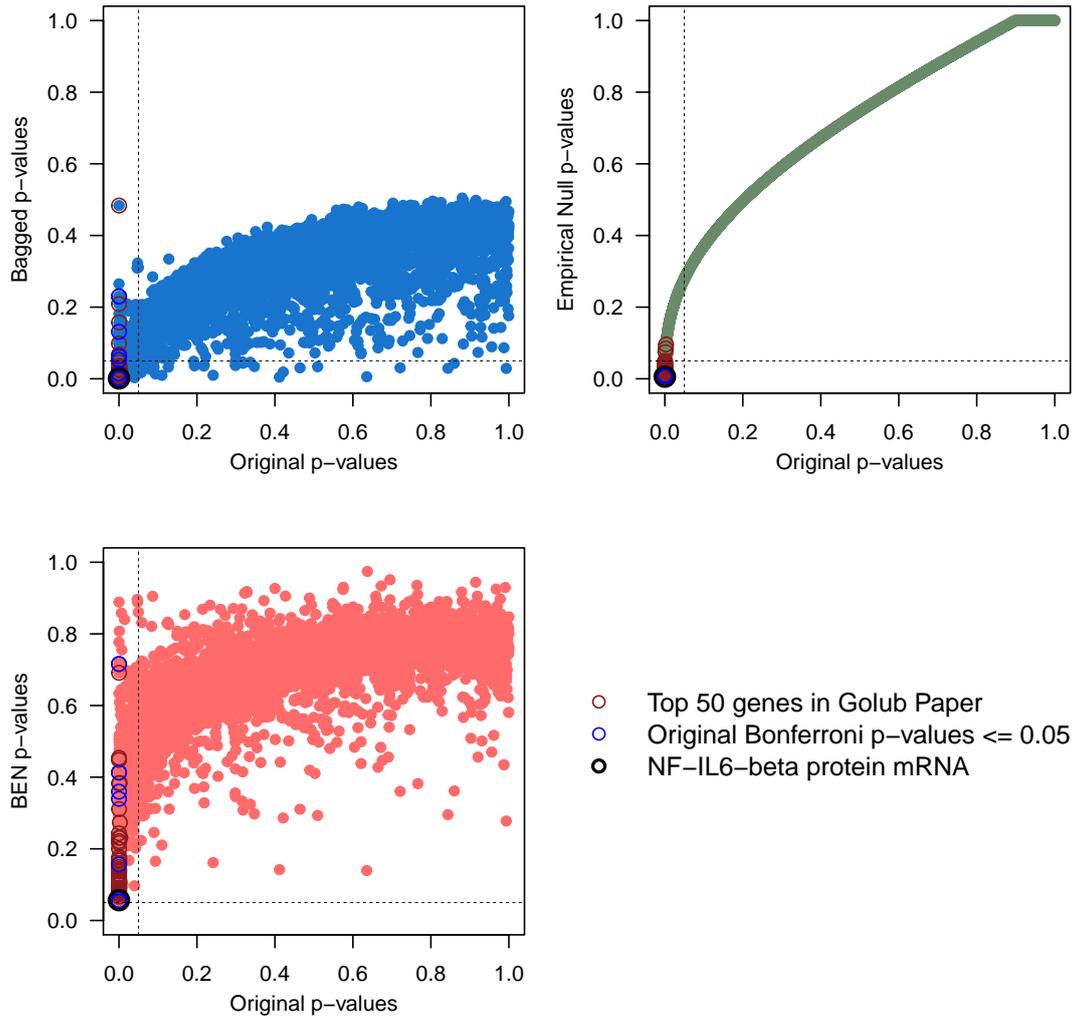

Fig. 2. The original *p*-values from the univariate logistic regressions are plotted against the Bagged *p*-values, Empirical $N(0.13, 1.70)$ Null recalibrated *p*-values, and the Bagged Empirical Null (BEN) *p*-values. The open red circles are the 50 *p*-values Golub *and others* (1999) reported as differentially expressed, and the open blue circles are the Bonferroni adjusted *p*-values $\leqslant 0.05$ from the original univariate logistic regressions. The bold black open circle is the top gene selected by the two-dimensional *p*-value and AUC criterion of the BEN algorithm, which corresponds the NF-IL6-beta protein mRNA gene.

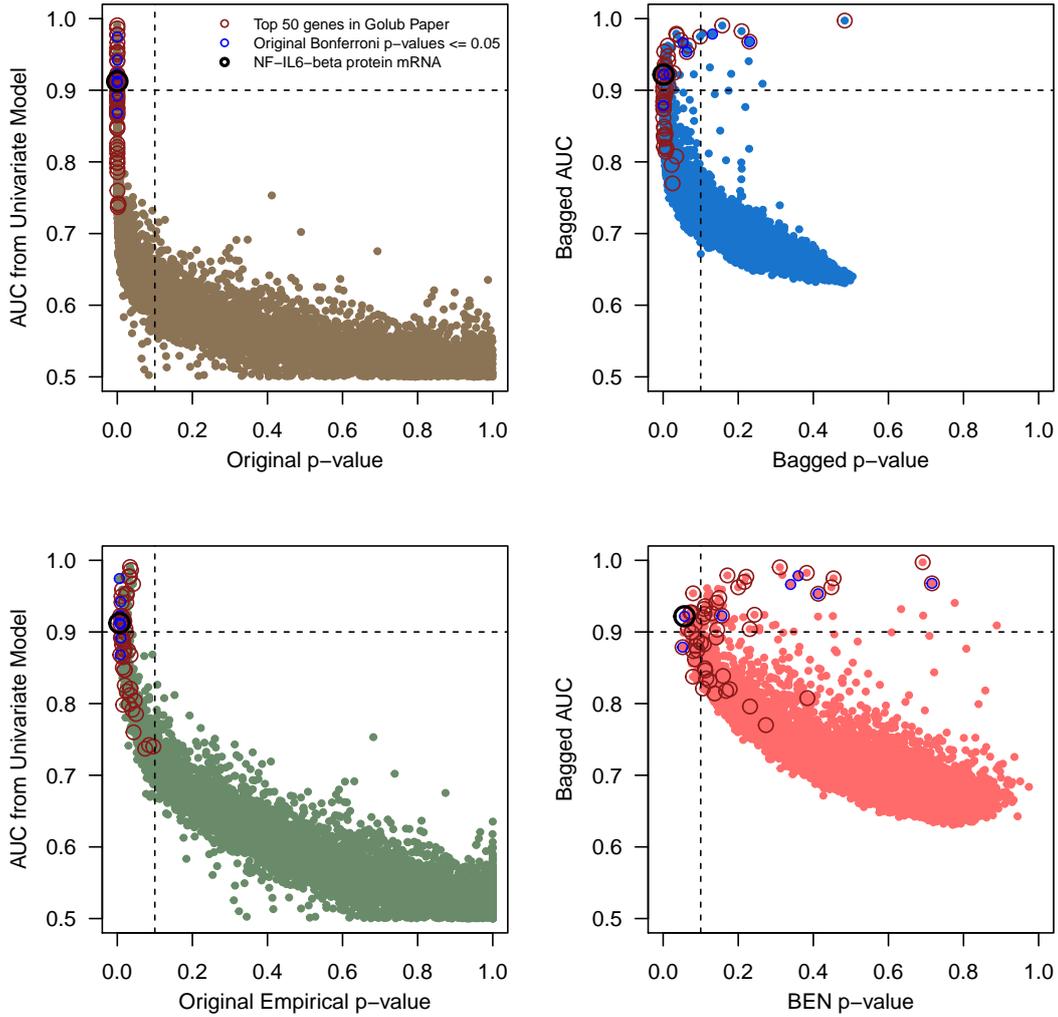

Fig. 3. The original *p*-values from the univariate logistic regressions and Empirical $N(0.13, 1.70)$ Null recalibrated *p*-values, are plotted against the corresponding model AUC. The Bagged *p*-values, and the Bagged Empirical Null (BEN) *p*-values are plotted against the Bagged AUC. The open red circles are the 50 *p*-values Golub *and others* (1999) reported as differentially expressed, and the open blue circles are the Bonferroni adjusted *p*-values $\leqslant 0.05$ from the original univariate logistic regressions. The bold black open circle is the top gene selected by the two-dimensional *p*-value and AUC criterion of the BEN algorithm, which corresponds the NF-IL6-beta protein mRNA gene.

## 5. Discussion

We have developed a Bagged Empirical Null strategy for obtaining $p$-values that do not need to be adjusted post-hoc when performing large scale inference. Based on our simulations and examples, combining these $p$-values with bagged model fit statistics appears to be advantageous as it tends to select truly differentially expressed genes more often than traditional $p$-value corrections. This direct approach to $p$-value calculations may be more useful when the goal is to identify the maximum number of truly significant genes, while controlling the FDR and maximizing the power.

Our motivation comes from gene expression microarray data analysis, where the same set of genes does not seem to be reproducible across different experiments of statistical methodologies. In this unique setting, it is highly impractical to pre-specify a different model for every gene, and so for situations where a large number of tests are to be evaluated (as in genetic and other high-dimensional data), current methodology can benefit from bootstrap aggregating procedures.

We have applied our method to pseudo-simulated gene expression data and the original leukemia data, and have demonstrated that our method is superior by having the most desirable Type I/Type II error tradeoff. A strength of our approach is the ability to consider any set of models (parameters, link function, non-parametric model) during bagging, as well as not being confined to the conventional theoretical null as the testing distribution. By incorporating bootstrapping, model selection, and empirical null procedures, the BEN algorithm has the advantage of using multi-dimensional gene selection metrics, beyond the single adjusted $p$-value traditionally used. Consequently, the BEN algorithm will lead to more robust and reproducible biological findings.

## 6. Description of Supplementary Materials

Supplementary Materials including technical appendices, additional results of the BEN algorithm using linear models when applied to the leukemia data, and annotated R scripts are available by email.

## Acknowledgments

The authors wish to thank Professors Robert Greevy, Matt Shotwell, Thomas Stewart, Melinda Aldrich, and Nathaniel Mercaldo for critical reading, helpful suggestions, and valuable feedback of the original version of the paper. In addition, the authors would like to thank Eric Grogan, and the TREAT Laboratory via funding from the Department of Veterans Affairs Career Development Award (10-024), and Melinda Aldrich via funding from NIH/NCI 5K07CA172294, as well as the Department of Thoracic Surgery for supporting this work. *Conflict of Interest*: None declared.